\documentclass{llncs}

\usepackage{amsmath}
\usepackage{graphics}
\usepackage{graphicx}
\usepackage{amssymb}
\usepackage{algorithmic}
\usepackage{algorithm}
\usepackage{boxedminipage}
\usepackage{booktabs}

\def\punto{\hspace*{\fill}\Box}

\newcommand{\squishlist}{
   \begin{list}{$\bullet$}
    { \setlength{\itemsep}{0pt}      \setlength{\parsep}{1.75pt}
      \setlength{\topsep}{3pt}       \setlength{\partopsep}{0pt}
      \setlength{\leftmargin}{1.5em} \setlength{\labelwidth}{1em}
      \setlength{\labelsep}{0.5em} } }

\newcommand{\squishend}{
    \end{list}  }

\newcommand{\set}[2]{\ensuremath{\{$ $#1$ $|$ $#2$ $\}}}
\newcommand{\setone}[1]{\ensuremath{\{#1\}}}

\begin{document}
\pagestyle{empty}

\title{Stop the Chase}

\author{Michael Meier\thanks{The work of this author was funded by DFG grant GRK 806/3.}, Michael Schmidt$^{\star}$ and Georg Lausen
}

\institute{University of Freiburg,
Institute for Computer Science \\
Georges-K\"ohler-Allee,
79110 Freiburg, Germany  \\
{\email{\{meierm,mschmidt,lausen\}@informatik.uni-freiburg.de}}
}

\maketitle

\begin{abstract}
The chase procedure, an algorithm proposed 25+ years ago to fix
constraint violations in database instances, has been successfully applied
in a variety of contexts, such as query optimization, data exchange, 
and data integration. Its practicability, however, is limited by the
fact that -- for an arbitrary set of constraints -- it might not terminate;
even worse, chase termination is an undecidable problem in general. In
response, the database community has proposed sufficient
restrictions on top of the constraints that guarantee chase termination 
on any database instance. In this paper, we propose a novel sufficient
termination condition, called {\it inductive restriction}, which strictly
generalizes previous conditions, but can be checked as efficiently.
Furthermore, we motivate and study the problem of {\it data-dependent} chase
termination and, as a key result, present sufficient termination conditions
w.r.t.~fixed instances. They are strictly more general
than inductive restriction and might guarantee termination
although the chase does not terminate in the general case.
\end{abstract}

\section{Introduction}

The chase procedure is a fundamental algorithm that has been successfully
applied in a variety of database
applications~\cite{mms1979,jk1982,bv1984,h2001,l2002,fkmp2005,dpt2006,ohk2009}.
Originally proposed to tackle the implication problem for data
dependencies~\cite{mms1979,bv1984} and to optimize Conjunctive
Queries (CQs) under data dependencies~\cite{asu1979,jk1982}, it has become
a central tool in Semantic Query Optimization (SQO)~\cite{pt1999,dpt2006,sml2008}.
For instance, the chase can be used to enumerate minimal CQs under a set of
dependencies~\cite{dpt2006}, thus supporting the search for more efficient
query evaluation plans. Beyond SQO, it has been applied in many other contexts, such as
data exchange~\cite{fkmp2005}, data integration~\cite{l2002}, query
answering using views~\cite{h2001}, and probabilistic databases~\cite{ohk2009}.

The core idea of the chase algorithm is simple: given a set of dependencies
(also called constraints) over a database schema and an instance as input,
it fixes constraint violations in the instance. One problem with the chase,
however, is that -- given an arbitrary set of constraints -- it might never
terminate; even worse, this problem is undecidable in general, also for
a fixed instance~\cite{dnr2008}. Addressing this issue, sufficient conditions for
the constraints that guarantee termination on any database instance have been
proposed~\cite{fkmp2005,dnr2008,sml2008}. Such conditions are the central topic
in this paper. In particular, we make two key contributions.

{\bf A novel sufficient termination condition for the chase.}
We introduce the class of {\it inductively restricted constraints}, for which
the chase terminates in polynomial time data complexity.
Like existent sufficient termination conditions, inductive restriction asserts
that there are no positions in the schema where fresh labeled nulls
might be cyclically created during chase application. It relies on a
sophisticated study of~(a)~positions in the database schema where null
values might appear, (b)~subsets of the constraints that cyclically pass
null values, and (c)~connections between such cycles. The combination of
these aspects makes inductive restriction more general than previous
sufficient termination conditions, thus making a larger class of constraints
amenable to the chase procedure.

{\bf Data-dependent chase termination.}
Whenever inductive restriction does not apply to a constraint set, no termination
guarantees for the general case can be derived. Arguably, reasonable applications
should never risk non-termination, so the chase algorithm cannot be safely applied
to {\it any instance} in this case. Tackling this problem, we study {\it data-dependent
chase termination:} given constraint set $\Sigma$ and a fixed instance~$I$,
does the chase with $\Sigma$ terminate on~$I$? This setting particularly makes sense
in the context of SQO, where the query -- interpreted as database instance -- is chased:
typically, the size of the query is small, so the ``data'' part can be analyzed efficiently
(as opposed to the case where the input is a large database instance).
We propose two complemental approaches.

Our first, static scheme relies on the observation that, when the instance $I$ is fixed, we
can safely ignore constraints in the constraint set that will never fire when chasing~$I$,
i.e.~if general sufficient termination conditions hold for those constraints that
might fire on $I$. As a fundamental result, we show that in general
it is undecidable if a constraint will never fire when chasing a fixed instance.
Nevertheless, we provide a sufficient condition that allows us to identify
such constraints, and derive a sufficient data-dependent termination condition.

Whenever this static approach fails, our second, dynamic approach comes into play:
we run the chase and track cyclically created fresh null values 
in a so-called monitor graph. We then fix the maximum depth of cycles in the monitor graph
and stop the chase when this limit is exceeded: in such a case, no termination guarantees
can be made. However, the search depth implicitly defines a class of constraint-instance
pairs for which the chase terminates. It can be seen as a natural condition
that allows us to stop the chase when ``dangerous'' situations arise. Hence,
our approach adheres to situations that might well cause non-termination and is
preferable to blindly running the chase and aborting after a fixed amount
of time, or a fixed number of chase steps. Applications might choose the
maximum search depth following a pay-as-you-go paradigm. Ultimately, the
combination of static and dynamic analysis allows us to safely apply the chase,
although no data-independent termination guarantees can be made. 

{\bf Structure.} We start with some preliminaries in the following section
and, in Section~\ref{sec:motivation}, continue with a discussion of
non-termination and a motivating example for data-dependent chase termination.
Section~\ref{sec:i} introduces inductive restriction, our sufficient (data-independent)
termination condition. Finally, we present our static and dynamic
approach to data-dependent chase termination in Section~\ref{sec:d}.

\section{Preliminaries}

{\bf General mathematical notation.} The natural numbers $\mathbb{N}$ do not include $0$. 
For $n \in \mathbb{N}$, we denote by $[n]$ the set $\setone{1,...,n}$. For a set $M$,
we denote by $2^M$ its powerset. Given a tuple $t = (t_1,\dots,t_n)$ we define the
tuple obtained by projecting on positions $1 \leq i_1 < \dots < i_m \leq n$ as
$p_{i_1,\dots,i_m}(t) := (t_{i_1},\dots,t_{i_m})$.

\textbf{Databases.} We fix three pairwise disjoint infinite sets:
the set of \textit{constants}~$\Delta$, the set of \textit{labeled nulls}
$\Delta_{null}$, and the set of \textit{variables} $V$. A \textit{database schema}
$\mathcal{R}$ is a finite set of relational symbols $\setone{R_1,...,R_n}$. 
In the rest of the paper, we
assume the database schema and the set of constants and labeled nulls to
be fixed. A \textit{database instance} $I$ is a finite set of $\mathcal{R}$-atoms that contains only elements from 
$\Delta \cup \Delta_{null}$ in its positions. We denote an element of an instance as \textit{fact}. The domain of $I$, $dom(I)$, is the set of elements from $\Delta \cup \Delta_{null}$ that appear in $I$.


We use the term \textit{position} to denote a position in a predicate, e.g.~a three-ary
predicate $R$ has three positions $R^1, R^2, R^3$. We say that a variable, labeled
null, or constant $c$ appears e.g. in a position $R^1$ if there exists a fact $R(c,...)$.

\textbf{Constraints.} Let $\overline{x}, \overline{y}$ be tuples of variables.
We consider two types of database constraints: \textit{tuple generating dependencies} 
(TGDs) and \textit{equality generating dependencies} (EGDs). A TGD is a first-order sentence 
$\alpha := \forall \overline{x} (\phi(\overline{x}) \rightarrow \exists \overline{y} \psi(\overline{x},\overline{y}))$ such that~(a) both $\phi$ and $\psi$ are conjunctions of atomic formulas (possibly with parameters from $\Delta$), (b) $\psi$ is not empty, (c) $\phi$ is possibly empty, (d) both $\phi$ and
$\psi$ do not contain equality atoms and (e) all variables from $\overline{x}$ that occur
in $\psi$ must also occur in $\phi$. We denote by $pos(\alpha)$ the set of positions in $\phi$. An EGD is a first-order sentence $\alpha := \forall \overline{x} (\phi(\overline{x}) \rightarrow x_i = x_j)$,
 where $x_i, x_j$ occur in $\phi$ and $\phi$ is a non-empty conjunction of equality-free
$\mathcal{R}$-atoms (possibly with parameters from $\Delta$). We denote by $pos(\alpha)$
the set of positions in $\phi$.
As a notational convenience, we will often omit the $\forall$-quantifier and
respective list of universally quantified variables. For a set of TGDs and EGDs
$\Sigma$ we set $pos(\Sigma) := \bigcup_{\xi \in \Sigma} pos(\xi)$.


\textbf{Chase.}
We assume that the reader is familiar with the chase procedure and give only a short
introduction here, referring the interested reader to~\cite{fkmp2005}
for a more detailed discussion. A chase step $I \stackrel{\alpha, \overline{a}}{\rightarrow} J$ takes a relational database
instance $I$ such that $I \nvDash \alpha(\overline{a})$ and adds tuples (in case of TGDs)
or collapses some elements (in case of EGDs) such that the resulting relational database
$J$ is a model of $\alpha(\overline{a})$. If $J$ was obtained from $I$ in that kind,
we sometimes also write $I\overline{a} \oplus C_{\alpha}$ instead of $J$. 
A chase sequence is an exhaustive application of applicable constraints 
$I_0 \stackrel{\alpha_0, \overline{a}_0}{\longrightarrow} I_1 \stackrel{\alpha_1, \overline{a}_1}{\longrightarrow} \ldots$,
where we impose no strict order on what constraint to apply in case several
constraints are applicable. If this sequence is finite, say $I_r$ being its final element,
the chase terminates and its result $I_0^{\Sigma}$ is defined as $I_r$. The length of this
chase sequence is $r$. Note that different orders of application orders may lead to a
different chase result. However, as proven in \cite{fkmp2005}, two different chase orders
always lead to homomorphically equivalent results, if these exist. Therefore, we write
$I^{\Sigma}$ for the result of the chase on an instance $I$ under constraints $\Sigma$.
It has been shown in~\cite{mms1979,bv1984,jk1982} that $I^{\Sigma} \models \Sigma$.
If a chase step cannot be performed (e.g., because application of an EGD would have to
equate two constants) or in case of an infinite chase sequence, the result of the
chase is undefined.
 

\section{A Motivating Example}
\vspace{-0.2cm}
\label{sec:motivation}

\begin{figure}[t]
\begin{boxedminipage}{12cm}
\underline{\bf Sample Schema:} \verb!hasAirport!({\it c\_id}),
\verb!fly!({\it c\_id1},{\it c\_id2},{\it dist}),
\verb!rail!({\it c\_id1},{\it c\_id2},{\it dist})\\
\underline{\bf Constraint Set:} $\Sigma := \setone{\alpha_1,\alpha_2,\alpha_3}$, where
\vspace{-0.25cm}
\begin{tabbing}
xxx \= \kill
$\alpha_1:$ If there is a flight connection between two cities, both of them 
have an airport:\\
\>\verb!fly!($x_1$,$x_2$,$y$) $\rightarrow$ \verb!hasAirport!($x_1$), \verb!hasAirport!($x_2$)\\

$\alpha_2:$ Rail-connections are symmetrical: \verb!rail!($x_1$,$x_2$,$y$) $\rightarrow$ \verb!rail!($x_2$,$x_1$,$y$)\\

$\alpha_3:$ Each city that is reachable via plane has at least one outgoing flight scheduled:\\
\>\verb!fly!($x_1$,$x_2$,$y_1$) $\rightarrow \exists$ $x_3$, $y_2$ \verb!flight!($x_2$,$x_3$,$y_2$)
\end{tabbing}
\end{boxedminipage}
\vspace{-0.4cm}
\caption{Sample Database Schema and Constraints of a Travel Agency.}
\label{fig:schemaandconstraints}
\vspace{-0.2cm}
\end{figure}

Non-termination of the chase is caused by fresh labeled null values
that are repeatedly created when fixing constraint violations. As an example,
consider the travel agency database in Figure~\ref{fig:schemaandconstraints}.
Predicate \verb!hasAirport!
contains cities that have an airport and \verb!fly! (\verb!rail!) stores
flight (rail) connections between cities, including their distance.
In addition to the schema, constraints $\alpha_1$-$\alpha_3$ have been specified,
e.g.~$\alpha_3$ might have been added to
assert that, for each city reachable via plane, the schedule is integrated in
the local database. Now consider the CQ $q_1$ below (in datalog notation,
with constant $c_1$ and variables $x_1$, $x_2$, $y_1$, $y_2$).

\vspace{-0.05cm}
\begin{tabbing}
x \= xxx \= \kill
\>$q_1$:\>\verb!rf!($x_2$) :- \verb!rail!($c_1$,$x_1$,$y_1$), \verb!fly!($x_1$,$x_2$,$y_2$)
\end{tabbing}
\vspace{-0.05cm}

The query selects all cities that can be reached from $c_1$ through
rail-and-fly. To chase $q_1$, we interpret its body as instance
I := \{\verb!rail!($c_1$,$x_1$,$y_1$),\verb!fly!($x_1$,$x_2$,$y_2$)\}, where 
$c_1$ is a constant and the $x_i$, $y_i$ labeled nulls. We observe that
$\alpha_3$ does not hold on $I$, since there is a flight to
city $x_2$, but no outgoing flight from $x_2$. To fix this violation,
the chase adds a new tuple $t_1 :=$ \verb!fly!($x_2$,$x_3$,$y_3$) to $I$,
where $x_3$, $y_3$ are fresh labeled null values. However, in the resulting instance
$I' := I \cup \{ t_1 \}$, $\alpha_3$ is again violated (this time for $x_3$)
and in subsequent steps the chase adds
\verb!fly!($x_3$,$x_4$,$y_4$), \verb!fly!($x_4$,$x_5$,$y_5$),
\verb!fly!($x_5$,$x_6$,$y_6$), $\dots$. Clearly, it will never terminate.

Reasonable applications should not risk non-termination, so for the constraint
set in Figure~\ref{fig:schemaandconstraints} termination is in
question for {\it all queries}, although there might be queries for which the
chase terminates. Tackling this problem, we propose to investigate data-dependent
chase termination, i.e.~to study sufficient termination guarantees
for a {\it fixed instance} when no general termination guarantees apply.
We illustrate the benefits of having such guarantees for query
$q_2$ below, which selects all cities $x_2$ that can be reached from $c_1$ via
rail-and-fly and the same transport route leads back from $x_2$ to $c_1$
($c_1$ is a constant, $x_i$, $y_i$ are variables).

\vspace{-0.05cm}
\begin{tabbing}
x \= xxx \= \kill
\>$q_2$:\>\verb!rffr!($x_2$) :- \verb!rail!($c_1$,$x_1$,$y_1$), \verb!fly!($x_1$,$x_2$,$y_2$), \verb!fly!($x_2$,$x_1$,$y_2$), \verb!rail!($x_1$,$c_1$,$y_1$)
\end{tabbing}
\vspace{-0.05cm}

Query $q_2$ violates only $\alpha_1$. The chase terminates and transforms $q_2$
into $q_2'$:

\vspace{-0.05cm}
\begin{tabbing}
x \= xxx \= xxxxxxxxx \= \kill
\>$q_2'$:\>\verb!rffr!($x_2$) :- \verb!rail!($c_1$,$x_1$,$y_1$), \verb!fly!($x_1$,$x_2$,$y_2$), \verb!fly!($x_2$,$x_1$,$y_2$), \verb!rail!($x_1$,$c_1$,$y_1$),\\
\>\>\>\verb!hasAirport!($x_1$), \verb!hasAirport!($x_2$)
\end{tabbing}
\vspace{-0.05cm}

The resulting query $q_2'$ satisfies all constraints and is a so-called 
{\it universal plan}~\cite{dpt2006}: intuitively, it incorporates all possible
ways to answer the query. As discussed in~\cite{dpt2006}, the universal plan
forms the basis for finding smaller equivalent queries (under the respective
constraints), by choosing subqueries of $q_2'$ and testing if they can be
chased to a query that is homomorphical to $q_2'$. Using this technique we can easily
show that the following two queries are equivalent to $q_2$.

\vspace{-0.05cm}
\begin{tabbing}
x \= xxx \= xxxxxxxxxxxx \= \kill
\>$q_2''$:\>\verb!rffr!($x_2$) :- \verb!rail!($c_1$,$x_1$,$y_1$), \verb!fly!($x_1$,$x_2$,$y_2$), \verb!fly!($x_2$,$x_1$,$y_2$)\\
\>$q_2'''$:\>\verb!rffr!($x_2$) :- \verb!hasAirport!($x_1$), \verb!rail!($c_1$,$x_1$,$y_1$), \verb!fly!($x_1$,$x_2$,$y_2$), \verb!fly!($x_2$,$x_1$,$y_2$)
\end{tabbing}
\vspace{-0.05cm}

Instead of $q_2$ we thus could evaluate $q_2''$ or $q_2'''$,
which might well be more performant: in both $q_2''$ and $q_2'''$
the join with \verb!rail!($x_1$,$c_1$,$y_1$) has been eliminated; moreover,
if \verb!hasAirport! is duplicate-free,  the additional join
of \verb!rail! with \verb!hasAirport! in $q_2'''$ may serve as a filter that
decreases the size of intermediate results and speeds up query evaluation.
This strategy is called {\it join introduction} in SQO (cf.~\cite{k1986}).
Ultimately, the chase for $q_2$ made it possible to detect 
$q_2''$ and $q_2'''$, so it would be desirable to have data-dependent termination
guarantees that allow us to chase~$q_2$ (and $q_2''$, $q_2'''$). We will present
such conditions in Section~\ref{sec:d}.

\vspace{-0.4cm}
\section{Data-independent Chase Termination}
\vspace{-0.2cm}
\label{sec:i}

In the past, sufficient conditions for constraint sets have been developed
that guarantee chase termination for any instance. One such
condition is {\it weak acyclicity}~\cite{fkmp2005}, which asserts
that there are no cyclically connected positions in the constraint set that may
introduce fresh labeled null values, by a global study of relations between the
constraints. In~\cite{dnr2008}, weak acyclicity was
generalized to {\it stratification}, which enforces weak acyclicity
only locally, for subsets of constraints that might cyclically cause to fire
each other. We further generalized stratification to {\it safe restriction} in~\cite{sml2008}.
We start by reviewing its central ideas and formal definition, which form the
basis for our novel condition {\it inductive restriction}.

{\bf Safe Restriction.} The idea of safe restriction is to keep track of positions
where fresh null values might be created in or copied to. As a basic tool, we borrow
the definition of {\it affected positions} from~\cite{cgk2008}. We emphasize that,
in~\cite{cgk2008}, this definition has been used in a different context: there,
the constraints are interpreted as axioms that are used to derive new facts from
the database and the problem is query answering on the implied database, using the
chase as a central tool.

\begin{definition} \cite{cgk2008} \em
Let $\Sigma$ be a set of TGDs. The set of \textit{affected positions}
$\mbox{aff}(\Sigma)$ is defined inductively as follows.
Let $\pi$ be a position in the head of an $\alpha \in$~$\Sigma$. 
\squishlist
	\item If an existentially quantified variable appears in $\pi$, then $\pi \in \mbox{aff}(\Sigma)$.
	\item If the same universally quantified variable $X$ appears both in position $\pi$, and only in affected positions in the body of $\alpha$, then $\pi \in \mbox{aff}(\Sigma)$. $\punto$
\squishend
\end{definition}

Akin to the dependency graph in weak acyclicity~\cite{fkmp2005}, we define a safety
condition that asserts the absence of cycles through constraints that may
introduce fresh null values. As an improvement, we exhibit the observation that only values
created due to or copied from affected positions may cause non-termination.
We introduce the notion of {\it propagation graph}, which refines the
dependency graph from~\cite{fkmp2005} by taking affected positions into consideration.

\begin{definition} \em
Let $\Sigma$ be a set of TGDs. We define a directed graph called
{\it propagation graph} $\mbox{prop}(\Sigma):=(\mbox{aff}(\Sigma),E)$ as follows.
There are two kinds of edges in $E$. Add them as follows: for every TGD 
$\forall \overline{x}(\phi(\overline{x}) \rightarrow \exists \overline{y} \psi(\overline{x},\overline{y})) \in \Sigma$
 and for every $x$ in $\overline{x}$ that occurs in $\psi$ and every occurrence of $x$ in $\phi$ in position $\pi_1$
\squishlist
	\item if $x$ occurs only in affected positions in $\phi$ then, for every occurrence of $x$ in $\psi$ in position $\pi_2$, add an edge $\pi_1 \rightarrow \pi_2$ (if it does not already exist).
	\item if $x$ occurs only in affected positions in $\phi$ then, for every existentially quantified variable $y$ and for every occurrence of $y$ in a position $\pi_2$, add a special edge $\pi_1 \stackrel{*}{\rightarrow} \pi_2$ (if it does not already exist). $\punto$
\squishend
\end{definition}

\begin{definition} \em
A set $\Sigma$ of constraints is called \textit{safe} iff $\mbox{prop}(\Sigma)$ has no cycles
going through a special edge. $\punto$
\end{definition}

Safety is a sufficient termination condition which~strictly generalizes weak acyclicity
and is different from stratification~\cite{sml2008}. The idea behind safe restriction
now is to assert safety locally, for subsets of the constraints that may cyclically cause
each other to fire in such a way that null values are passed in these cycles. 

\begin{definition} \label{def-verf} \em
Let $\Sigma$ abe given and $P \subseteq pos(\Sigma)$. For all $\alpha, \beta \in \Sigma$, we define $\alpha \prec_{P} \beta$ iff there are tuples $\overline{a}, \overline{b}$ and a database instance $I$ s.t. (i) $I \nvDash \alpha(\overline{a})$, (ii) $I \models \beta(\overline{b})$, (iii) $I \stackrel{\alpha, \overline{a}}{\rightarrow} J$, (iv) $J \nvDash \beta(\overline{b})$,
(v) $I$ contains null values only in positions from $P$ and (vi) there is a null value $n \in \overline{b} \cap \Delta_{null}$ in the head of $\beta(\overline{b})$. $\punto$
\end{definition}

Informally, $\alpha \prec_P \beta$ holds if $\alpha$ might cause $\beta$ to fire
s.t., when null values occur only in positions from P, $\beta$
copies some null values. We next introduce a notion for affected positions
relative to a constraint and a set of positions. 

\begin{definition} \em
For any set of positions $P$ and a TGD $\alpha$ let $\mbox{aff-cl}(\alpha,P)$ be the set
of positions $\pi$ from the head of $\alpha$ such that

\squishlist
	\item for every universally quantified variable $x$ in $\pi$: $x$ occurs in the body of $\alpha$ only in positions from $P$ or
	\item $\pi$ contains an existentially quantified variable. $\punto$
\squishend
\end{definition}

On top of previous definitions we introduce the central tool of {\it restriction systems}.

\begin{definition} \label{rest} \em
A {\it restriction system} is a pair $(G'(\Sigma),f)$, where
$G'(\Sigma) := (\Sigma,E)$ is a directed graph and
$f: \Sigma \rightarrow 2^{pos(\Sigma)}$ is a function such that
\squishlist
	\item forall TGDs $\alpha$ and forall $(\alpha,\beta) \in E$: $\mbox{aff-cl}(\alpha,f(\alpha))  \cap pos(\setone{\beta}) \subseteq f(\beta)$,
		\item forall EGDs $\alpha$ and forall $(\alpha,\beta) \in E$: $f(\alpha)  \cap pos(\setone{\beta}) \subseteq f(\beta)$, and
	\item forall $\alpha, \beta \in \Sigma$: $\alpha \prec_{f(\alpha)} \beta \implies (\alpha,\beta) \in E$. 
\squishend

A restriction system is {\it minimal} if it is obtained from
($(\Sigma,\emptyset)$,\{$(\alpha,\emptyset)$ $\mid$ $\alpha \in \Sigma$\}) by
a repeated application of the constraints from bullets one to three (until all
constraints hold) s.t., in case of the first and second bullet, the image of
$f(\beta)$ is extended only by those positions that are required to satisfy
the condition.$\punto$
\end{definition}

\begin{example} 
Let predicate E($x$,$y$) store graph edges and predicate S($x$) store some nodes.
The constraints $\Sigma = \{ \alpha_1, \alpha_2 \}$ with
$\alpha_1 :=$ S($x$), E($x$,$y$) $\rightarrow$ E($y$,$x$) and
$\alpha_2 :=$ S($x$), E($x$,$y$) $\rightarrow$ $\exists z$ E($y$,$z$), E($z$,$x$)
assert that all nodes in S have a cycle of length $1$ and $2$. It holds that
$\mbox{aff}(\Sigma)$ = \{E$^1$,E$^2$\} and it is easy to verify that~$\Sigma$
is neither safe nor stratified (see~Def.~2 in~\cite{dnr2008}). The minimal
restriction system for $\Sigma$ is
G'($\Sigma$):=($\Sigma$,\{($\alpha_2$,$\alpha_1$)\}) with
f($\alpha_1$) := \{E$^1$,E$^2$\} and f($\alpha_2$) := $\emptyset$; in particular, 
$\alpha_1 \not \prec_{f(\alpha_1)} \alpha_1$,
$\alpha_1 \not \prec_{f(\alpha_1)} \alpha_2$,
$\alpha_2 \prec_{f(\alpha_2)} \alpha_1$, and
$\alpha_2 \not \prec_{f(\alpha_2)} \alpha_2$ hold.
$\punto$
\label{ex:sr1}
\end{example}

As shown in~\cite{sml2008}, the minimal restriction system is unique and can be
computed by an~\textsc{NP}-algorithm. We are ready to define the notion of
safe restriction:

\begin{definition} \label{def-rest} \em
$\Sigma$ is called {\it safely restricted} if and only if every strongly
connected component of its minimal restriction system is safe. $\punto$
\end{definition}

\begin{example} Constraint set $\Sigma$ from Example~\ref{ex:sr1} is safely
restricted: its minimal restriction system contains no strongly connected
components.$\punto$
\label{ex:sr2}
\end{example}

As shown in~\cite{sml2008}, safe restriction (a)~guarantees chase termination
in polynomial time data complexity, (b)~is strictly more general than stratification,
and (c)~it can be checked by a $\textsc{coNP}$-algorithm if a set of constraints
is safely restricted.

{\bf Inductive Restriction.} We now introduce the novel class of {\it inductively
restricted constraints}, which generalizes safe restriction but, like the latter,
gives
polynomial-time termination guarantees. We start with a motivating example.

\begin{example}
We extend the constraints from Example~\ref{ex:sr1} to
$\Sigma' := \Sigma \cup \{ \alpha_3 \}$, where $\alpha_3 := \exists x, y S(x), E(x,y)$.
Then G'($\Sigma'$):=($\Sigma'$,\{($\alpha_1,\alpha_2$),($\alpha_2$,$\alpha_1$),($\alpha_3$,$\alpha_1$),($\alpha_3$,$\alpha_2$)\})
with f($\alpha_1$) = f($\alpha_2$) := \{E$^1$,E$^2$,S$^1$\} and
f($\alpha_3$) := $\emptyset$ is the minimal restriction system. It contains the strongly
connected component \{$\alpha_1$,$\alpha_2$\}, which is not safe. 
Consequently, $\Sigma'$ is not safely restricted.$\punto$
\label{ex:ir1}
\end{example}

\begin{figure}[t]
\centering
\begin{boxedminipage}{10cm}
\begin{tabbing}
l \= lllll \= lllllll \= lllllllll \= \kill
{\bf part}($\Sigma$: Set of TDGs and EGDs) \{\\
\>1:\>compute the strongly connected components (as sets of constraints) $C_1$, $\dots$, $C_n$\\
\>\> of the minimal restriction system of $\Sigma$;\\
\>2:\>$D \leftarrow \emptyset$\\
\>3:\>{\bf if}\ \  (n == 1)\ \  {\bf then}\ \ \\
\>4:\>\> {\bf if}\ \ ($C_1 \not = \Sigma$)\ \ {\bf then}\ \ return {\bf part}($C_1$); {\bf endif}\\
\>7:\>\> return $\setone{\Sigma}$;\\
\>8:\> {\bf endif}\\\>6:\>{\bf for}\ \  i=1 to n\ \  {\bf do}\ \ $D \leftarrow D\ \cup$ {\bf part}($C_i$);\ \ {\bf endfor}\\
\>11:\>return $D$; \}
\end{tabbing}
\end{boxedminipage}
\vspace{-0.3cm}
\caption{Algorithm to compute subsets of $\Sigma$.}
\vspace{-0.4cm}
\label{fig:algopart}
\end{figure}

Intuitively, safe restriction does not apply in the example above because
$\alpha_3$ ``infects'' position S$^1$ in the restriction system.
Though, null values cannot be repeatedly created in S$^1$: $\alpha_3$
fires at most once, so it does not affect chase termination. Our novel
termination condition recognizes such situations by recursively computing
the minimal restriction systems of the strongly connected components. We
formalize this computation in Algorithm~1, called {\it part}($\Sigma$). Based
on this algorithm, we define an improved sufficient termination condition.

\begin{definition} \em
Let $\Sigma$ be a set of constraints. We call $\Sigma$ {\it inductively
restricted} iff
for all $\Sigma' \in {\it part}(\Sigma)$ it holds that $\Sigma'$ is safe. $\punto$
\end{definition}

As stated in the following lemma, inductive restriction strictly generalizes safe restriction,
but does not increase the complexity of the recognition problem. 

\begin{lemma} \label{relindsafe} \em Let $\Sigma$ be a set of constraints.
\squishlist
\item If $\Sigma$ is safely restricted, then it is inductively restricted.
\item There is some $\Sigma$ that is inductively restricted, but not safely restricted.
\item The recognition problem for inductive restriction is in $\textsc{coNP}$. $\punto$
\squishend
\label{lemma:ir}
\end{lemma}

\begin{example} \label{bulletzwo}
Consider $\Sigma'$ from Example~\ref{ex:ir1}. It is easy to verify that
${\it part}(\Sigma') = \emptyset$ and we conclude that $\Sigma'$ is inductively
restricted. As argued in Example~\ref{ex:ir1}, $\Sigma'$~is not safely restricted,
which proves the second claim in Lemma~\ref{lemma:ir}.$\punto$
\label{ex:ir2}
\end{example}

The next theorem gives the main result of this section, showing that
inductive restriction guarantees chase termination in polynomial time data complexity.
To the best of our knowledge inductive restriction is the most general
sufficient termination condition for the chase that has been proposed so far.

\begin{theorem} \label{ind-main} \em
Let $\Sigma$ be a fixed set of inductively restricted constraints. Then, there exists a
polynomial $Q \in \mathbb{N}[X]$ such that for any database instance $I$, the length
of every chase sequence is bounded by $Q(||I||)$, where $||I||$ is the number of distinct
values in $I$. $\punto$
\end{theorem}

\section{Data-dependent Chase Termination}
\vspace{-0.15cm}
\label{sec:d}

{\bf Static Termination Guarantees.} 
Motivated by the example in Section~\ref{sec:motivation}, we now study data-dependent
chase termination: given a constraint set $\Sigma$ and a fixed instance $I$, does
the chase with $\Sigma$ terminate on $I$? Our first, static scheme relies on the observation
that the chase will always terminate on instance $I$ if the subset of constraints
that might fire when chasing $I$ with $\Sigma$ is inductively restricted. We call a constraint
$\alpha \in \Sigma$ {\it $(I,\Sigma)$-irrelevant} iff there is no chase sequence
$I \stackrel{\alpha_1, \overline{a_1}}{\longrightarrow} \dots \stackrel{\alpha,\overline{a}}{\longrightarrow} \dots$
and formalize our observation in Lemma~\ref{lem:relterm} below.

\begin{lemma}\em \label{lem:relterm}
Let $\Sigma' \subseteq \Sigma$ s.t. $\Sigma \setminus \Sigma'$ is a set of
$(I,\Sigma)$-irrelevant constraints. If $\Sigma'$ is inductively restricted, then the chase with
$\Sigma$ terminates for instance $I$.$\punto$
\end{lemma}

Hence, the crucial point is to effectively compute $(I,\Sigma)$-irrelevant constraints.
Unfortunately, one can show that $(I,\Sigma)$-irrelevance is undecidable in general.

\begin{theorem} \em
Let $\Sigma$ be a set of constraints, $\alpha \in \Sigma$ a constraint,
and $I$ an instance. It is undecidable if $\alpha$ is $(I,\Sigma)$-irrelevant.$\punto$
\label{th:irr}
\end{theorem} 

This result prevents us from computing the minimal set of constraints that will
fire when chasing $I$. Still, we can give sufficient conditions that guarantee
$(I,\Sigma)$-irrelevance for a constraint. We specify such a condition
on top of the {\it chase graph} introduced in~\cite{dnr2008}. The chase graph for $\Sigma$ is
the graph $G(\Sigma) =(\Sigma,\prec)$, where $\alpha \prec \beta$ holds for
$\alpha, \beta \in \Sigma$ iff the first three bullets from Def.~\ref{def-verf}
hold. It was shown in \cite{dnr2008} that, given $\Sigma$, the chase graph can be
computed by an $\textsc{NP}$-algorithm.

\begin{proposition} \label{prop:bla} \em Let $I$ be an instance and $\Sigma$ be a set of constraints. Further
let $\alpha_I := \exists \overline{x} \bigwedge_{R(\overline{x}') \in I}  R(\overline{x}')$
where $\overline{x} := \bigcup_{R(\overline{x}') \in I} \overline{x}'$.
If the chase graph $G(\Sigma \cup \setone{\alpha_I})$ contains no directed path from 
 $\alpha_I$ to $\beta \in \Sigma$, then $\beta$ is $(I,\Sigma)$-irrelevant.$\punto$
\end{proposition}

Proposition~\ref{prop:bla} combined with Lemma~\ref{lem:relterm} gives us a
sufficient data-dependent condition for chase termination, as illustrated in the
following example.
 
\begin{example}
Consider constraint set $\Sigma$ from Fig.~\ref{fig:schemaandconstraints} and 
$q_2$ from Section~\ref{sec:motivation}. We set
$\alpha_I$:=$\exists$ $c_1$,$x_1$,$x_2$,$y_1$,$y_2$ \verb!rail!($c_1$,$x_1$,$y_1$), \verb!fly!($x_1$,$x_2$,$y_2$), \verb!fly!($x_2$,$x_1$,$y_2$), \verb!rail!($x_1$,$c_1$,$y_1$)
and compute the chase graph
$G(\Sigma \cup \setone{\alpha_I}) := (\Sigma \cup \setone{\alpha_I}, \{(\alpha_I,\alpha_1), (\alpha_3,\alpha_3)\})$.
By Proposition~\ref{prop:bla}, $\alpha_2$ and $\alpha_3$ are $(I,\Sigma)$-irrelevant. It holds
that $\Sigma \setminus \setone{\alpha_2,\alpha_3} = \setone{\alpha_1}$
is inductively restricted, so we know from Lemma~\ref{lem:relterm} that the
chase of $q_2$ with $\Sigma$ terminates. Similar argumentations hold for
$q_2''$ and $q_2'''$ from Section~\ref{sec:motivation}.$\punto$
\end{example}
 
{\bf Monitoring Chase Execution.} 
If the previous data-dependent termination condition does not apply, we propose
to monitor the chase run and abort if tuples are created that may potentially
lead to non-termination. We introduce a data structure called {\it monitor graph}
that allows us to track the chase run.

\vspace{-0.05cm}
\begin{definition} \em
A {\it monitor graph} is a tuple $(V,E)$, where $V \subseteq \Delta_{null} \times 2^{\mbox{pos}(\Sigma)}$ and $E \subseteq V \times \Sigma \times 2^{\mbox{pos}(\Sigma)} \times V$. $\punto$
\end{definition}

A node in a monitor graph is a tuple $(n,\pi)$, where $n$ is a database value and
$\pi$ the positions in which $n$ was first created (e.g. as null value
with the help of some TGD). An edge $(n_1,\pi_1,\varphi_i,\Pi,n_2,\pi_2)$ between
$(n_1,\pi_1)$, $(n_2,\pi_2)$ is labeled with the constraint $\varphi_i$
that created $n_2$ and the set of positions $\Pi$ from the body of $\varphi_i$
in which $n_1$ occurred when $n_2$ was created. The monitor graph is successively
constructed while running the chase, according to the following definition.

\vspace{-0.05cm}
\begin{definition} \em
The monitor graph $G_{\mathcal{S}}$ w.r.t. 
$\mathcal{S} = I_0 \stackrel{\varphi_0, \overline{a}_0}{\longrightarrow} 
\ldots \stackrel{\varphi_{r-1}, \overline{a}_{r-1}}{\longrightarrow} 
I_r$ is a monitor graph that is inductively defined as follows
\squishlist
	\item $G_0 = (\emptyset,\emptyset)$ is the empty chase segment graph.
	\item If $i < r$ and $\varphi_i$ is an EGD then $G_{i+1} := G_i$.
	\item If $i < r$ and $\varphi_i$ is a TGD then $G_{i+1}$ is obtained from $G_i = (V_i,E_i)$ as follows. If the chase step $I_i \stackrel{\varphi_i, \overline{a}_i}{\longrightarrow} I_{i+1}$  does not introduce any new null values, then $G_{i+1} := G_{i}$. Otherwise, $V_{i+1}$ is set as the union of $V_i$ and all pairs $(n,\pi)$, where $n$ is a newly introduced null value and $\pi$ the set of positions in which $n$ occurs. $E_{i+1} := E_i \cup \set{(n_1,\pi_1,\varphi_i,\Pi,n_2,\pi_2)}{(n_1,\pi_1) \in V_i, (n_2,\pi_2) \in V_{i+1} \backslash V_i \text{ and } \Pi \text{ is the set of positions in } body(\varphi_i(\overline{a}_i)) \text{ where } n_1 \text{ occurs}}$. $\punto$
\squishend
\end{definition}

Our next task is to define a necessary criterion for non-termination on top of the
monitor graph. To this end, we introduce the notion of {\it k-cyclicity}.

\begin{definition} \em
Let $G = (V,E)$ be a monitor graph and $k \in \mathbb{N}$. $G$ is called $k$-cyclic if and only if there are pairwise distinct $v_1,...,v_k \in V$ such that 
\squishlist
\item there is a path in $E$ that sequentially contains $v_1$ to $v_k$ and
\item for all $i \in [k-1]$: $p_{2,3,4,6}(v_i) = p_{2,3,4,6}(v_{i+1})$.$\punto$
\squishend
\end{definition}

We call a chase  sequence {\it $k$-cyclic} if its monitor graph is $k$-cyclic.
A chase sequence may potentially be infinite if some finite prefix is $k$-cyclic,
for any $k \geq 1$: 

\begin{lemma} \label{dynamic} \em
Let $k \in \mathbb{N}$. If there is some infinite chase sequence $\mathcal{S}$ when chasing $I_0$ with $\Sigma$, then there is some finite prefix  of $\mathcal{S}$ that is $k$-cyclic. $\punto$
\end{lemma}

To avoid non-termination, an application can fix a cycle-depth $k$ and stop the chase when
this limit is exceeded. For every terminating chase sequence there is a
$k$ s.t.~the sequence is not $k$-cyclic, so if $k$ is chosen large enough the chase
will succeed. We argue that $k$-cyclicity is a {\it natural} condition that considers
only situations that may cause non-termination, so our approach it is preferable to blindly chasing the
instance and stopping after a fixed amount of time or number of chase steps.
As justified by the following proposition, the choice of $k$ follows a pay-as-you-go principle:
for larger $k$-values the chase will succeed in more cases.
We refer the interested reader to the proof of the proposition for an example.

\begin{proposition} \em \label{prop:nutzendynamic}
For each $k \in \mathbb{N}$ there is some $\Sigma_k$ and $I_k$ s.t.~(a)~both $\Sigma_k$ and
the subset of constraints in $\Sigma_k$ that are not $(I_k,\Sigma_k)$-irrelevant are not inductively
restricted; (b)~every chase sequence for $I_k$ with $\Sigma_k$ is $(k-1)$-, but not $k$-cyclic.$\punto$
 \end{proposition}


{
\small
\bibliographystyle{abbrv}
\bibliography{main}

\begin{thebibliography}{10}

\bibitem{asu1979}
A.~V. Aho, Y.~Sagiv, and J.~D. Ullman.
\newblock Efficient {O}ptimization of a {C}lass of {R}elational {E}xpressions.
\newblock {\em ACM Trans. Database Syst.}, 4(4):435--454, 1979.

\bibitem{bv1984}
C.~Beeri and M.~Y. Vardi.
\newblock A {P}roof {P}rocedure for {D}ata {D}ependencies.
\newblock {\em J. ACM}, 31(4):718--741, 1984.

\bibitem{cgk2008}
A.~Cal\`{\i}, G.~Gottlob, and M.~Kifer.
\newblock Taming the {I}nfinite {C}hase: {Q}uery {A}nswering under {E}xpressive
  {R}elational {C}onstraints.
\newblock In {\em Descr. Logics}, volume 353, 2008.

\bibitem{dnr2008}
A.~Deutsch, A.~Nash, and J.~Remmel.
\newblock The {C}hase {R}evisited.
\newblock In {\em PODS}, pages 149--158, 2008.

\bibitem{dpt2006}
A.~Deutsch, L.~Popa, and V.~Tannen.
\newblock Query {R}eformulation with {C}onstraints.
\newblock {\em S{I}{G}{M}{O}{D} {R}ecord}, 35(1):65--73, 2006.

\bibitem{h2001}
A.~Y. Halevy.
\newblock Answering {Q}ueries {U}sing {V}iews: {A} {S}urvey.
\newblock {\em VLDB Journal}, pages 270--294, 2001.

\bibitem{jk1982}
D.~S. Johnson and A.~Klug.
\newblock Testing {C}ontainment of {C}onjunctive {Q}ueries under {F}unctional
  and {I}nclusion {D}ependencies.
\newblock In {\em PODS}, pages 164--169, 1982.

\bibitem{k1986}
J.~J. King.
\newblock Q{U}{I}{S}{T}: {A} {S}ystem for {S}emantic {Q}uery {O}ptimization in
  {R}elational {D}atabases.
\newblock In {\em VLDB}, pages 510--517, 1981.

\bibitem{l2002}
M.~Lenzerini.
\newblock Data {I}ntegration: {A} {T}heoretical {P}erspective.
\newblock In {\em PODS}, pages 233--246, 2002.

\bibitem{mms1979}
D.~Maier, A.~Mendelzon, and Y.~Sagiv.
\newblock Testing {I}mplications of {D}ata {D}ependencies.
\newblock In {\em SIGMOD}, pages 152--152, 1979.

\bibitem{ohk2009}
D.~Olteanu, J.~Huang, and C.~Koch.
\newblock S{P}{R}{O}{U}{T}: {L}azy vs. {E}ager {Q}uery {P}lans for
  {T}uple-{I}ndependent {P}robabilistic {D}atabases.
\newblock In {\em ICDE}, 2009.
\newblock To appear.

\bibitem{pt1999}
L.~Popa and V.~Tannen.
\newblock An {E}quational {C}hase for {P}ath-{C}onjunctive {Q}ueries,
  {C}onstraints, and {V}iews.
\newblock In {\em ICDT}, pages 39--57, 1999.

\bibitem{fkmp2005}
{R. Fagin et al.}
\newblock Data {E}xchange: {S}emantics and {Q}uery {A}nswering.
\newblock {\em Theor. Comput. Sci.}, 336(1):89--124, 2005.

\bibitem{sml2008}
M.~Schmidt, M.~Meier, and G.~Lausen.
\newblock Foundations of {S}{P}{A}{R}{Q}{L} {Q}uery {O}ptimization.
\newblock {\em CoRR}, abs/0812.3788, 2008.

\end{thebibliography}
}

\newpage
\appendix
{\large \bf APPENDIX}
\section{Proof of Lemma~\ref{relindsafe}}
\begin{enumerate}
	\item Let $\Sigma$ be safely restricted. By construction every element $\Sigma' \in {\it part}(\Sigma)$ is contained in some strongly connected component $C_{\Sigma'}$ of the minimal restriction system of $\Sigma$. By assumption $C_{\Sigma'}$ is safe, so every subset of $C_{\Sigma'}$ is also safe. Thus, $\Sigma'$ is safe.
	
	\item See Example~\ref{bulletzwo}.
	
	\item It was shown in \cite{sml2008} that the relation $\prec_P$ (for a set of positions $P$) can be decided by an $\textsc{NP}$-algorithm. For an input $\Sigma$, the set ${\it part}(\Sigma)$ can thus be computed in non-deterministic polynomial time. To check whether $\Sigma$ is not inductively restricted, guess some $\Sigma' \in {\it part}(\Sigma)$ and verify that it is not safe. This implies our claim.
\end{enumerate}

%

\section{Proof of Theorem~\ref{ind-main}}
The proof of this theorem is by induction on the depth of the recursive calls $n$ during the execution of algorithm {\it part} with input $\Sigma$. If $n=0$, $\Sigma$ is safely restricted and it was shown in \cite{sml2008} that the chase terminates in polynomial time data complexity for this case. If $n > 0$, then consider the strongly connected components $C_1,...,C_n$ of the minimal restriction system of $\Sigma$.
By induction hypothesis, chasing with $C_i$ terminates in time $Q_i(||I||)$.
The rest of this proof is analogous to the construction in the induction step
from the proof of Theorem~11 in~\cite{sml2008} (showing that the chase for
safely restricted constraints terminates in polynomial-time data complexity)
and therefore is omitted here.

\section{Proof of Lemma~\ref{lem:relterm}}
It holds that $\Sigma'$ contains all constraints that may fire during the execution of the chase starting with $I$ and $\Sigma$. So, if $I^{\Sigma'}$ exists then $I^{\Sigma}$ exists and $I^{\Sigma'}=I^{\Sigma}$. If $\Sigma'$ is inductively restricted, then $I^{\Sigma'}$ exists, which implies the claim.

\section{Proof of Theorem~\ref{th:irr}}
It is well-known that the following problem is undecidable: given a Turing machine $M$ and and a state transition $t$ from the description of $M$, does $M$ reach $t$ (given the empty string as input)? From $(M,t)$, we will compute a set of TGDs and EGDs $\Sigma_M$ and a TGD $\alpha_t \in \Sigma_M$ such that the following equivalence holds: $M$ reaches $t$ (given the empty string as input) $\Leftrightarrow$ there is a chase sequence in the computation of the chase with $\Sigma_M$ applied to the empty instance such that $\alpha_t$ will eventually fire.

Our reduction uses the construction in the proof of Theorem~1 in~\cite{dnr2008}.  To be self-contained, we review it here again. We use the signature consisting of the relation symbols: $T(x, a, y)$ tape ``horizontal'' edge from $x$ to $y$ with symbol $a$; $H(x, s, y)$ head ``horizontal'' edge from $x$ to $y$ with state $s$; $L(x, y)$ left ``vertical'' edge; $R(x, y)$ right ``vertical'' edge; 
$A_{\delta}(x),B_{\delta}(x)$ for every stater transition $\delta$, one constant for every tape symbol, one constant for every head state, the special constant $B$ marking the beginning of the tape and $\square$ to denote an empty tape cell. The set of constraints $\Sigma_M$ is as follows.

\begin{enumerate}
	\item To set the initial configuration:\\ $\exists w, x, y, z T(w,B, x), T (x,\square, y),H(x, s_0, y),T(y,E,z)$\\
where $\square$ is the blank symbol and $s_0$ is the initial state (both are constants).

\item For every state transition $\delta$ which moves the head to the right,
replacing symbol $a$ with $a'$ and going from state $s$ to state $s'$:\\
$T(x, a, y),H(x, s, y), T (y, b, z) \rightarrow$\\
$\exists x', y', z' L(x, x'), R(y, y'), R(z, z'), T(x', a', y'),$\\
 $T(y', b, z'),H(y', s', z'),A_{\delta}(w')$.\\
 Here $a, s, a', b$, and $s'$ are constants.

\item For every state transition $\delta$ which moves the head to the right
past the end of the tape replacing symbol a with a' and going
from state s to state s':\\
$T(x, a, y),H(x, s, y), T (y, E, z) \rightarrow$\\
$\exists w', x', y', z' L(x, x'), R(y, y'), R(z, z'), T(x', a', y'),$\\
 $T(y', \square, z'),H(y', s', z'), T(y',E,w'),A_{\delta}(w')$.\\
 Here $a, s, a', b$, and $s'$ are constants.

\item Similarly for state transitions which move the head to the left.


\item Similarly for state transitions which do not move the head.

\item For every state transition $\delta$:\\
$A_{\delta}(x) \rightarrow B_{\delta}(x)$


\item Left copy:\\
$T(x, a, y), L(y, y') \rightarrow \exists x' L(x, x'), T(x', a, y')$.\\
 Here $a$ is a  constant.

\item Right copy:\\
$T(x, a, y),R(x, x') \rightarrow \exists y' T(x', a, y'),R(y, y')$.\\
 Here $a$ is a  constant.
\end{enumerate}

The state transition $t$ is transformed to $\alpha_t$ in the same way like in bullet six above. It is crucial to the proof that every state transition $\delta$ in $M$ is represented as a single TGD $A_{\delta}(x) \rightarrow B_{\delta}(x)$. The constraint for the initial configuration fires exactly once. The computation of the chase with this set of constraint can be understood as a grid and each row in the grid represents a configuration of the Turing machine. It can be shown that $(M,t)$ is a yes-instance if and only if $(\Sigma_M,\alpha_t)$ is a yes-instance. Thus, the equivalence from above holds.

\section{Proof of Proposition~\ref{prop:bla}}
Assume that $\beta$ is not $(I,\Sigma)$-irrelevant. Then, there is a chase sequence $I \stackrel{\alpha_1, \overline{a_1}}{\longrightarrow} I_1 \stackrel{\alpha_2, \overline{a_2}}{\longrightarrow} \dots \stackrel{\alpha_r, \overline{a_r}}{\longrightarrow} I_r \stackrel{\beta,\overline{a}}{\longrightarrow} \dots$.
If $\alpha_I \prec \beta$ we are finished. Otherwise, there must be some $n_r \in [r]$ such that $\alpha_{n_r} \prec \beta$ (otherwise $\beta$ could not fire). If $\alpha_I \prec \alpha_{n_r}$ we are finished. Otherwise, there must be some $n_{r-1} \in [n_r - 1]$ such that $\alpha_{n_{r-1}} \prec \alpha_{n_r}$ (otherwise $\alpha_{n_r}$ could not fire). After some finite amount of iterations of this process we have that $\alpha_I \prec \alpha_{n_1} \prec ... \prec \alpha_{n_r} \prec \beta$. Therefore, the chase graph contains a directed path from $\alpha_I$ to $\beta$.

\section{Proof of Lemma~\ref{dynamic}}
Assume that 
\begin{itemize}
 \item we have an infinite chase sequence $\mathcal{S} = (I_i)_{i \in \mathbb{N}}$ and 
 \item there is some $k \in \mathbb{N}$ such that every finite prefix of $\mathcal{S}$ is not $k$-cyclic.
\end{itemize}

Let $(\mathcal{S}_i)_{i \in \mathbb{N}}$ be the sequence of finite prefixes of $\mathcal{S}$ (such that $\mathcal{S}_i$ is a chase sequence of length $i$) and let $(G_{\mathcal{S}_i})_{i \in \mathbb{N}}$ the respective sequence of monitor graphs. A path in a monitor graph is a finite sequence of edges $e_1,...,e_l$ (and not of nodes) such that $p_{5,6}(e_i)=p_{1,2}(e_{i+1})$ for $i \in [l-1]$. 

We define the notion of {\it depth} of a node in a monitor graph. Let $v$ be a node in $G_{\mathcal{S}_i}$ and $pred(v)$ the set of predecessors of $v$. In case $v$ has no predecessors, the depth of $v$, $depth_{G_{\mathcal{S}_i}}(v)$, is defined as zero. In case $v$ has predecessors, then $depth_{G_{\mathcal{S}_i}}(v) := 1 + max \set{depth_{G_{\mathcal{S}_i}}(w)}{w \in pred(v)}$.\\

The following claim follows immediately from the definition of the monitor
graph. The formal proof is left to the reader.

\begin{proposition}\em \label{basic}
Let $v$ be a node in $G_{\mathcal{S}_i}$ and $j > i$.

\squishlist
 \item $G_{\mathcal{S}_i}$ is an acyclic labeled tree.
 \item Every null value that appears in $I_i$ appears in some first position of a node in $G_{\mathcal{S}_i}$.
	\item There is a homomorphism\footnote{A homomorphism leaves relational symbols and constraints untouched, i.e.~is the identity on elements from $\Delta$.} $h_{ij}$ from $G_{\mathcal{S}_i}$ to $G_{\mathcal{S}_j}$ such that $depth_{G_{\mathcal{S}_i}}(v) \leq depth_{G_{\mathcal{S}_j}}(h_{ij}(v))$.
	 \item If $I_i \stackrel{\varphi_i, \overline{a}_i}{\rightarrow}I_{i+1}$, $b \in \overline{a}_i$ is a null value and $c$ a null value that was newly created in this step, then the depth of any node in $G_{\mathcal{S}_{i+1}}$ in which $b$ appears is strictly smaller than the depth of any node in $G_{\mathcal{S}_{i+1}}$ in which $c$ appears. (Proof by induction on $i$) $\punto$
\squishend
\end{proposition}
	 
	The next proposition is the most important step in the proof of this lemma and follows directly from bullet four in Proposition~\ref{basic}.
	 
	\begin{proposition}\em \label{adv} Let $i \in \mathbb{N}$. For every  $d \in \mathbb{N} \cup \setone{0}$ there is a number $k_d \in \mathbb{N}$ such that for every $i \in \mathbb{N}$ it holds that $|\set{v}{depth_{G_{\mathcal{S}_i}}(v) \leq d}| \leq k_d$. \textit{Note that $k_d$ is independent from $i$.} (Proof by induction on $d$) $\punto$
	 \end{proposition}

We observe another fact.
	\begin{proposition}\em
There is some $p_k \in \mathbb{N}$ such that if some $G_{\mathcal{S}_i}$ has a path of length $p_k$, then $\mathcal{S}_i$ is $k$-cyclic. $\punto$
\end{proposition}

This is because we have only a bounded number of relational symbols and constraints available. The remaining step in the proof is to show that if we choose~$i$ large enough, then $G_{\mathcal{S}_{i}}$ contains a path of length $p_k$. Assume that this claim does not hold. By Proposition~\ref{adv}, the number of nodes of a certain depth is bounded (independent of $i$). So, if for any $i$ there would be no path of length $p_k$ in $G_{\mathcal{S}_{i}}$, then the number of nodes in $G_{\mathcal{S}_{i}}$ would be bounded (independent of $i$). This implies that the chase has introduced only a bounded number of fresh null values, which contradicts the assumption of an infinite chase sequence.

\section{Proof of Proposition~\ref{prop:nutzendynamic}}

We set $I_k := \setone{S(c_1),...,S(c_k), R_k(c_1,...,c_k)}$ and\\
$\Sigma_k := \setone{\varphi}$, where $\varphi := S(x_k), R_k(x_1,...,x_k) \rightarrow \exists y R_k(y,x_1,...,x_{k-1})$.\\

First observe that $\Sigma_k$ contains no $(I,\Sigma_k)$-irrelevant constraints, so the subset
of the constraints in $\Sigma_k$ that is not $(I,\Sigma)$-irrelevant equals to $\Sigma_k$.
It is easy to verify that $\Sigma_k$ is not inductively restricted, although the chase with
$\Sigma_k$ always terminates, independent of the underlying data instance, so condition~(a) holds. 
 
We now chase of $I_k$ with $\Sigma_k$. There is only one possible chase sequence
$(J_i)_{0 \leq i \leq k}$, defined as $J_0 := I_k$, for
$i \leq k$: $J_{i} := J_{i-1} \cup \setone{R(n_i,...,n_1,c_1,...,c_{k-i})}$,
 and $n_1,...,n_k$ are fresh null values. It holds that $J_k \models \Sigma_k$.
 
 The monitor graph w.r.t. $(J_i)_{0 \leq i \leq k}$ is $(V,E)$, where $E := \set{(n_i,R_k^1)}{i \in [k]}$ and $V := \set{(n_i,R_k^1,\varphi,R_k^{j-i},n_j,R_k^1)}{1 \leq i < j \leq k}$.
We observe that the sequence is $(k-1)$-cyclic because $(n_1,R_k^1,\varphi,R_k^{1},n_2,R_k^1),...,(n_{k-1},R_k^1,\varphi,R_k^{1},n_k,R_k^1)$ constitute a path in the chase graph that satisfies the conditions of the definition of $(k-1)$-cyclicity. The chase sequence is not $k$-cyclic because there is no path of length at least $k$ in the monitor graph. This proves part~(b) of the proposition.

%
%
%
%

\end{document}